\documentclass{hcig}

\usepackage{amsmath}
\usepackage{graphicx}
\usepackage{hyperref}
\usepackage{algorithm}
\usepackage{algpseudocode}
\usepackage{amsfonts,amssymb}
\usepackage{mathtools}
\usepackage{bm}
\usepackage{wrapfig}
\usepackage{xcolor}
\usepackage[nameinlink]{cleveref}
\usepackage{array,tabularx,booktabs}

\newcolumntype{C}{>{\centering\arraybackslash}X}
\newcolumntype{W}{>{\centering\arraybackslash\hsize=1.2\hsize}X}
\newcolumntype{S}{>{\centering\arraybackslash\hsize=.9\hsize}X}

\algrenewcommand\algorithmicrequire{\textbf{Input:}}
\algrenewcommand\algorithmicensure{\textbf{Output:}}

\title{PSI3D: Plug-and-Play 3D Stochastic Inference with Slice-wise Latent Diffusion Prior}

\author{Wenhan Guo, Jinglun Yu, Yaning Wang, Jin U. Kang, and Yu Sun\textsuperscript{\Letter}}
\address{Johns Hopkins University\\\smallskip
{\footnotesize \textsuperscript{\Letter}Corresponding author: ysun214@jh.edu}}

\headertitle{PSI3D}
\headerauthors{Guo et al.}

\begin{document}

\maketitle
\thispagestyle{firstpagestyle}

\begin{abstract}
Diffusion models are highly expressive image priors for Bayesian inverse problems. However, most diffusion models cannot operate on large-scale, high-dimensional data due to high training and inference costs. In this work, we introduce a \textit{Plug-and-play algorithm for 3D stochastic inference with latent diffusion prior (PSI3D)} to address massive ($1024\times 1024\times 128$) volumes. Specifically, we formulate a Markov chain Monte Carlo approach to reconstruct each two-dimensional (2D) slice by sampling from a 2D latent diffusion model. To enhance inter-slice consistency, we also incorporate total variation (TV) regularization stochastically along the concatenation axis. We evaluate our performance on optical coherence tomography (OCT) super-resolution. Our method significantly improves reconstruction quality for large-scale scientific imaging compared to traditional and learning-based baselines, while providing robust and credible reconstructions.
\vspace{1.5mm}

The code for PSI3D is available at \href{https://github.com/Hopkins-CIG/PSI3D}{github.com/Hopkins-CIG/PSI3D}.
\end{abstract}

\section{Introduction}

Modern computational imaging problems are commonly modeled as the inverse problem
\begin{equation}
    \mathbf{y} = \mathbf{A}\mathbf{x} + \mathbf{e}, \qquad \mathbf{e}\sim \mathcal{N}(0,\sigma^2 \mathbf{I}), \label{eq:inv}
\end{equation}
where $\mathbf{x}\in\mathbb{R}^n$ is the unknown image (or volume), $\mathbf{y}\in\mathbb{R}^m$ is the measurement, and $\mathbf{A} \in \mathbb{R}^{m\times n}$ is the forward operator. We aim to recover the unknown $\mathbf{x}$ from the noisy measurement $\mathbf{y}$. Typical applications include computed tomography (CT), magnetic resonance imaging (MRI), microscopy, and astronomical imaging \cite{Barutcu:2021, Ahmad.etal2019, Tian.etal14, Akiyama.etal2019}. In many cases, the inverse problem is ill-posed, and image regularization becomes highly important for resolving fine structure from limited data and significant noise. A widely adopted strategy is to decouple data fidelity from prior modeling. 
\textit{Plug-and-play (PnP)} methods, for example, alternate between a data-consistency step and a prior step, where advanced denoisers are used as implicit priors \cite{Venkatakrishnan.etal2013, Chan.etal2016, Ryu.etal2019, Kamilov.etal2023}.

Diffusion models (DMs) have emerged as powerful and expressive image priors for PnP and other Bayesian algorithms \cite{ Chung.etal2023diffusion, Feng.etal2023scorebased, Sun.etal2024}. Recent works incorporate DMs into Bayesian solvers for inverse problems by viewing the prior step as sampling from the posterior of a Gaussian denoising problem. 
Specifically, \cite{Coeurdoux.etal2024, wu.etal2024pnpdm, Xu.etal2024provably} formulate a split Gibbs sampler that alternates between a likelihood step and a prior step, using diffusion models~\cite{karras2022edm} for sampling from the denoising posterior.
This enables a sampling-based optimization approach with reconstruction robustness crucial for scientific and medical imaging. 

\begin{figure}
    \centering
    \includegraphics[width=1.0\linewidth]{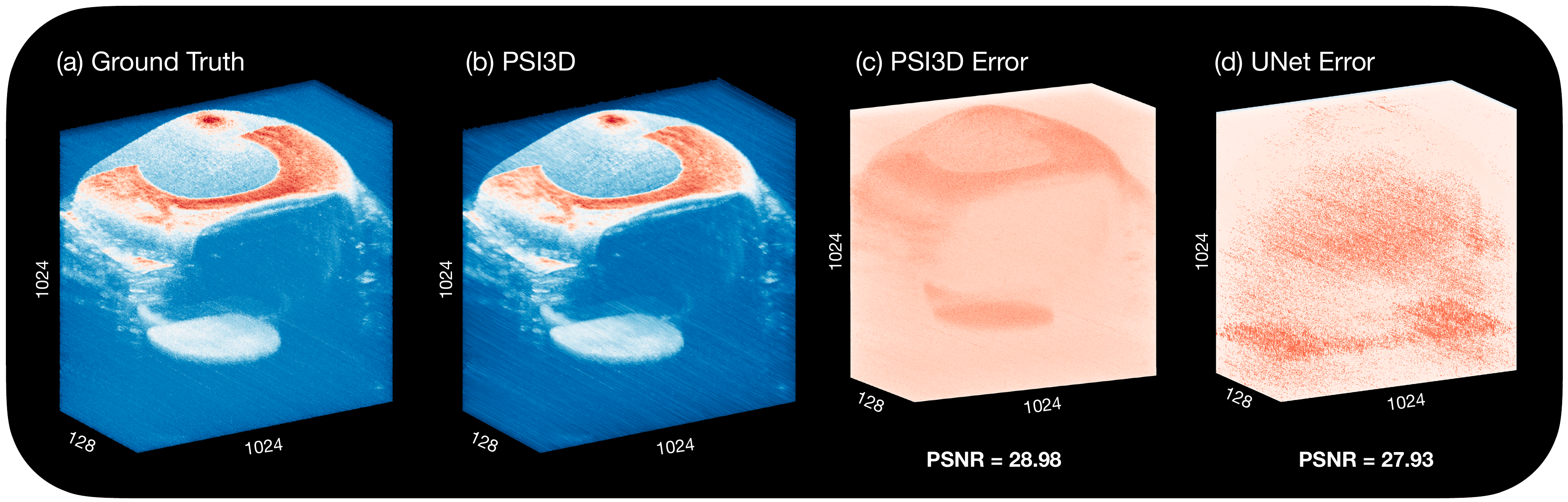}
    \caption{
    Visual comparison of the 3D OCT volumes ($1024 \times 1024 \times 128$) reconstructed by PSI3D and 3D UNet.
    From left to right, we show the ground truth volume, PSI3D reconstructed volume, absolute error for PSI3D, and absolute error for 3D UNet. Note that PSI3D accurately reconstructs the volume with minimal artifacts and a higher peak signal-to-noise ratio (PSNR).}
    \label{fig:volumes}
\end{figure}

However, scaling DM-based Bayesian solvers to large three-dimensional (3D) volumes remains challenging. On the one hand, training a 3D diffusion model is prohibitively expensive in terms of training data and compute. On the other hand, a naive slice-wise application of 2D diffusion priors often leads to insufficient regularization between slices. To resolve this, several recent methods combine pretrained 2D diffusion priors with inter-slice constraints. For example, DiffusionMBIR \cite{chung2023diffusionmbir} uses model-based iterative reconstruction to augment a 2D diffusion prior with a total variation (TV) prior imposed on the orthogonal axis. It then formulates an alternating direction method of multipliers update on the aggregated slices. However, DiffusionMBIR operates on moderate volumes $(256^3)$ and still faces practical limitations when targeting very large volumes. Other similar methods, including diffusion posterior sampling \cite{Chung.etal2023diffusion} and latent-variable optimization \cite{ozaki2024iterative}, show strong 3D performance but either operate on smaller spatial extents or produce a deterministic solution that limits uncertainty quantification. To the best of our knowledge, current DM-based solvers cannot handle reconstructions at the scale of hundreds of millions of voxels.

We propose PSI3D to extend the PnP-DM framework in \cite{wu.etal2024pnpdm} to massive (e.g. $1024 \times 1024 \times 128$) 3D volumes with principled sampling and UQ. For each 2D slice, we project the 2D image onto latent space via a vector-quantized autoencoder (VQGAN) \cite{esser2021taming}. We then train an EDM diffusion model on the latent code, which reduces memory and computation substantially compared to image-space diffusion. To enforce consistency between slices, we impose 1D TV regularization per batch along the concatenation axis. We select each batch of slices from the volume with randomized rounding and alteration \cite{raghavan1987randomizedrounding, bansal2012rrwithalteration}, which guarantees stochasticity and multi-coverage in a given number of batches. 

We evaluate PSI3D on optical coherence tomography (OCT) super-resolution. Compared to traditional regularization and recent learning-based baselines, PSI3D significantly improves reconstruction quality and provides posterior samples with credibility. We show an example volumetric reconstruction in Fig.~\ref{fig:volumes}. While we demonstrate on OCT super-resolution, our method is agnostic to the imaging modality and the forward model physics. Our results suggest that latent PnP-DM with TV is a powerful and robust solver for general inverse problems. The MACE framework also provides sufficient flexibility for other priors useful for specific imaging modalities.

\section{Preliminaries}

In this section, we introduce the preliminaries to our method, which focus on deterministic formulations. We consider the linear inverse problem in \eqref{eq:inv}. First write the negative log-posterior as an energy
\begin{equation}
    \mathcal{U}(\mathbf{x};\mathbf{y}) \;=\; 
    \underbrace{f(\mathbf{x};\mathbf{y})}_{\text{likelihood}} + \underbrace{g(\mathbf{x})}_{\text{prior}} \;=\; 
    \tfrac{1}{2\sigma^2}\|\mathbf{y}-\mathbf{A}\mathbf{x}\|_2^2
    \;+\;
    g(\mathbf{x}) .
    \label{eq:energy}
\end{equation}
It is generally difficult to directly access the posterior due to complications of the prior. One widely-used method is half-quadratic splitting (HQS) \cite{Geman.Yang1995} to decouple the likelihood from the prior. Consider the HQS formulation with penalty $\rho>0$ and an auxiliary variable $\mathbf{z}$, where we write each iteration into two proximal subproblems:
\begin{align}
    \mathbf{x}^{t+1} 
    &= \arg\min_{\mathbf{x}}\; \Big\{ \, f(\mathbf{x};\mathbf{y}) + \tfrac{1}{2 \rho^2}\|\mathbf{x}-\mathbf{z}^t\|_2^2 \, \Big\} ,
    \tag{M1}\label{eq:M1}\\
    \mathbf{z}^{t+1} 
    &= \arg\min_{\mathbf{z}}\; \Big\{ \, g(\mathbf{z}) + \tfrac{1}{2 \rho^2}\|\mathbf{z}-\mathbf{x}^{t+1}\|_2^2 \, \Big\} .
    \tag{M2}\label{eq:M2}
\end{align}
where $g(\mathbf{z}) = \Phi_{3\mathrm{D}}(\mathbf{z})$ is ideally a $3$D prior for a $3$D inverse problem. It is computationally prohibitive to directly operate a diffusion prior in three dimensions. Instead, we restructure the prior as 
\begin{equation}
    g(\mathbf{z}) 
    \;=\; \sum_{i} \phi(\mathbf{z}_i) \;+\; \sum_{i} \lambda\,\mathrm{TV}_z(\mathbf{z}),
    \label{eq:prior-2d-tv}
\end{equation}
where $\mathbf{z}_i\!\in\!\mathbb{R}^{H\times W}$ is the $i$-th axial slice, $\phi(\cdot)$ is the energy of the (implicit) 2D diffusion prior, and $\mathrm{TV}_z$ is first-order isotropic TV along the orthogonal axis. We substitute \eqref{eq:prior-2d-tv} into \eqref{eq:M2} to obtain a single proximal problem with slice-wise diffusion and depth-wise TV.

We update a contiguous batch $B_t$ of size $B$ at iteration $t$. For each batch of slices in the volume, we have step \eqref{eq:M2} as
\begin{equation}
    \mathbf{z}^{t+1}_{B_t} 
    \;=\; \arg\min_{\{\mathbf{z}_i\}_{i\in B_t}} \\
        \Big\{ \sum_{i\in B_t}\phi(\mathbf{z}_i) + \sum_{i\in B_t} \lambda\,\mathrm{TV}_z(\mathbf{z}_{B_t})
        + \tfrac{1}{2 \rho^2}\|\mathbf{z}_{B_t}-\mathbf{x}^{t+1}_{B_t}\|_2^2 \Big\},
    \label{eq:batch-prox}
\end{equation}
and the other slices remain unchanged, i.e. $\mathbf{z}^{t+1}_i=\mathbf{z}^t_i$ for $i\notin B_t$. Using proximal average \cite{Sun.etal2020}, we perform slice-wise update in a batch by writing the approximation
\begin{equation}
    \mathbf{z}^{t+1}_{B_t} 
    \;\approx\; \sum_{i\in B_t} \arg\min_{\{\mathbf{z}_i\}_{i\in B_t}} \\
        \Big\{ \phi(\mathbf{z}_i) + \lambda\,\mathrm{TV}_z(\mathbf{z}_{B_t})
        + \tfrac{1}{2 \rho^2}\|\mathbf{z}_{B_t}-\mathbf{x}^{t+1}_{B_t}\|_2^2 \Big\},
    \label{eq:proximal avg}
\end{equation}
where we approximate \eqref{eq:M2} by an average of small proximals, substantially reducing computational cost per iteration. In the next section, we draw analogies from proximal average to implement $\phi$ via a slice-wise 2D diffusion prior step, followed by a TV prior step to provide inter-slice consistency within each batch.

\begin{figure*}
    \centering
    \includegraphics[width=1\linewidth]{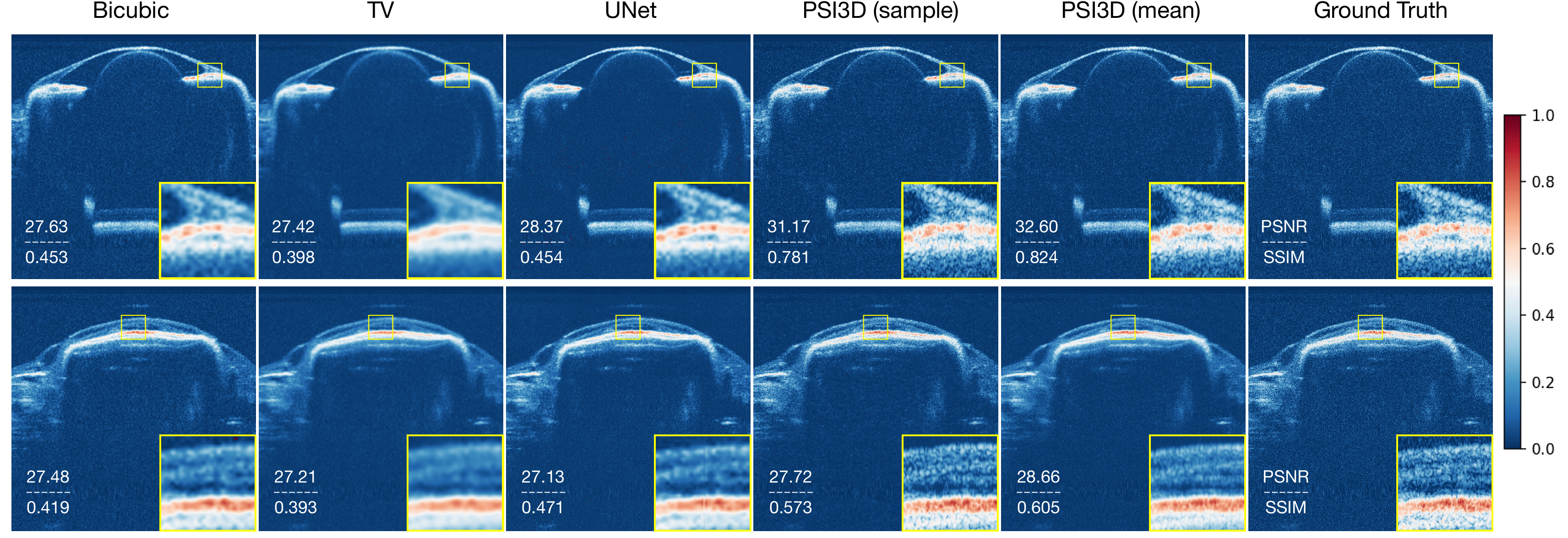}
    \caption{Visual comparison of slices from 3D reconstructions obtained using PSI3D and baseline methods.. Each row shows a single B-scan slice in the OCT volume, with a zoom-in view in the yellow boxes. Note that PSI3D accurately recovers fine anatomical details and achieves higher PSNR and SSIM than baseline methods.}
    \label{fig:slices}
\end{figure*}

\section{Method}

In this section, we introduce the PSI3D method. We use the split Gibbs sampler \cite{vono.dobigeon.chainais2019} to separate the likelihood and two complementary priors. First define
\[
g_{\text{d}}(\mathbf{z}) \triangleq -\log p_{\text{d}}(\mathbf{z}),\quad
g_{\text{tv}}(\mathbf{w}) \triangleq - \log p_{\text{TV}_z}(\mathbf{w}),
\]
where $p_{\text{d}}(\mathbf{z})$ and $p_{\text{TV}_z}(\mathbf{w})$ are the $2$D diffusion and $z$-directional TV prior distributions, respectively. We can then write the augmented posterior with two auxiliary variables $\mathbf{z}$, $\mathbf{w}$ and penalties $\rho_{\text{d}}$ and $\rho_{\text{tv}}$:
\begin{equation}
\pi(\mathbf{x},\mathbf{z},\mathbf{w}\mid\mathbf{y})
\;\propto\;
\exp\!\Big(-f(\mathbf{x};\mathbf{y}) - g_{\text{d}}(\mathbf{z}) - g_{\text{tv}}(\mathbf{w}) \\
-\tfrac{1}{2 \rho_{\text{d}}^2}\|\mathbf{z}-\mathbf{x}\|_2^2
-\tfrac{1}{2 \rho_{\text{tv}}^2}\|\mathbf{x}-\mathbf{w}\|_2^2
\Big).
\label{eq:augpost}
\end{equation}
We proceed to sample from \eqref{eq:augpost} by alternating between three conditional updates: the likelihood step, the diffusion prior step, and the TV prior step. 
\\
\\
\textbf{Likelihood Step:} sample $\mathbf{x}^{t+1}\!\sim\!p(\mathbf{x}\mid\mathbf{z}^t,\mathbf{w}^t,\mathbf{y})$, where
\begin{equation}
p(\mathbf{x}\mid\mathbf{z}^t,\mathbf{w}^t,\mathbf{y})
\;\propto\;
\exp\!\Big(-f(\mathbf{x};\mathbf{y}) \\
-\tfrac{1}{2 \rho_{\text{d}}^2}\|\mathbf{z}^t-\mathbf{x}\|_2^2
-\tfrac{1}{2 \rho_{\text{tv}}^2}\|\mathbf{x}-\mathbf{w}^t\|_2^2\Big).
\label{eq:likelihood step}
\end{equation}
For linear $\mathbf{A}$ and Gaussian noise with covariance $\sigma^2 \, \mathbf{I}$, this is equivalent to sampling from $\mathcal{N}\left(\boldsymbol{\mu}(\mathbf{z}^t, \mathbf{w}^t \right), \boldsymbol{\Lambda}^{-1})$, where
\[
\mathbf{\Lambda} = \mathbf{A}^\top\mathbf{A} /{\sigma^2} + \mathbf{I}/{\rho_\text{d}^2} + \mathbf{I}/{\rho_\text{tv}^2} \,,
\]
\[
\boldsymbol{\mu} (\mathbf{z}^t, \mathbf{w}^t)=\mathbf{\Lambda}^{-1} \left(\mathbf{A}^\top\mathbf{y}/{\sigma^2} + \mathbf{z}^t /{\rho_{\text{d}}^2} + \mathbf{w}^t / {\rho_{\text{tv}}^2} \right).
\]
However, computing the inverse $\boldsymbol{\Lambda}^{-1}$ is expensive. Following \cite{vono.dobigeon.chainais2022}, if there exists a singular value decomposition of $\mathbf{A} = \mathbf{U} \: \mathbf{S} \: \mathbf{V}^\top$, we can write the Cholesky factorization
\[
\boldsymbol{\Lambda}^{-1} = \mathbf{L} \mathbf{L}^\top, \quad
\mathbf{L} = \mathbf{V} \left( \mathbf{S}^2 /\sigma^2 + \mathbf{I}/{\rho_\text{d}^2} + \mathbf{I}/{\rho_\text{tv}^2} \right)^{-1/2},
\]
and form the sample $\mathbf{x} = \boldsymbol{\mu}(\mathbf{z}^t, \mathbf{w}^t) + \mathbf{L} \, \boldsymbol{\xi}_t$, where $\boldsymbol{\xi}_t \sim \mathcal{N}(0, \mathbf{I})$. For more general cases, we can sample from \eqref{eq:likelihood step} via gradient-based Monte Carlo methods such as Langevin dynamics \cite{Pereyra2016, Brose.etal2017, Song.etal2019}. 
\\
\\
\textbf{Diffusion Prior Step:} sample $\mathbf{z}^{t+1}\!\sim\!p(\mathbf{z}\mid\mathbf{x}^{t+1})$, where
\begin{equation}
p(\mathbf{z}\mid\mathbf{x}^{t+1})
\;\propto\;
\exp\!\Big(- g_{\text{d}}(\mathbf{z}) -\tfrac{1}{2 \rho_{\text{d}}^2}\|\mathbf{z}-\mathbf{x}^{t+1}\|_2^2 \Big) .
\label{eq:diffusion prior step}
\end{equation}
Importantly, this is the posterior of a Gaussian denoising problem with noisy ``measurement'' $\mathbf{x}^{t+1}$ and noise level $\rho_{\text{d}}$. In fact, \cite{wu.etal2024pnpdm} shows that we can directly sample from \eqref{eq:diffusion prior step} by running a reverse diffusion starting from a timestep with noise level $\rho_{\text{d}}$; see also \cite{Coeurdoux.etal2024, Xu.etal2024provably}.
Inspired by proximal average \cite{Sun.etal2020}, we approximate the proximal on the batch with slice-wise diffusion model sampling.
\\
\\
\textbf{TV Prior Step:} sample $\mathbf{w}^{t+1}\!\sim\!p(\mathbf{w}\mid\mathbf{x}^{t+1})$, where
\begin{equation}
p(\mathbf{w}\mid\mathbf{x}^{t+1})
\;\propto\;
\exp\!\Big(- g_{\text{tv}}(\mathbf{w}) -\tfrac{1}{2 \rho_{\text{tv}}^2}\|\mathbf{w}-\mathbf{x}^{t+1}\|_2^2 \Big).
\label{eq:tv prior step}
\end{equation}
There are many efficient dual-domain solvers for the proximal associated with \eqref{eq:tv prior step}, including Chambolle \cite{Chambolle.2004} and Beck-Teboulle \cite{Beck.Teboulle2009}, but they are deterministic and therefore do not yield samples. Following \cite{Bouman.etal2023generative}, we approximate a draw from the proximal generator by injecting AWGN $\boldsymbol{\eta}_t\!\sim\!\mathcal{N}(\mathbf{0},\mathbf{I})$ around a proximal point:
\begin{equation}
\mathbf{w}^{t+1}
\;=\;
\operatorname{prox}_{\rho_\text{tv}^{2}\, g_{\text{tv}}}\!\big(\mathbf{x}^{t+1}\big)
\;+\;
\rho_\text{tv}\,\boldsymbol{\eta}_t ,
\label{eq:gpnp}
\end{equation}
We apply \eqref{eq:gpnp} on a randomly chosen contiguous batch of slices to reduce memory and computational cost. 

Alg.~\ref{alg} presents the pseudocode for PSI3D. 
Another way to interpret the three stochastic updates is to view them as three agents in the multi-agent consensus equilibrium (MACE) framework \cite{Buzzard.etal2017, majee2021mace}. 
Iterative updates across the three steps conceptually form a probabilistic variant of MACE.
\\
\\
\textbf{Empirical Design Choices.} 
For the diffusion prior step, it is still computationally demanding to run a diffusion model on a $2$D batch with size $1024 \times 1024 \times B$. We implement \eqref{eq:diffusion prior step} in a latent space, where each $2$D slice ($1024 \times 1024$) is projected to a lower-dimensional latent $\mathbf{u}_i=\mathcal{E}(\mathbf{z}_i)$ via a VQGAN encoder $\mathcal{E}$ \cite{esser2021taming}. We run an EDM diffusion model in latent space to draw $\mathbf{u}^{t+1}_i\!\sim\!p(\mathbf{u}\mid\mathcal{E}(\mathbf{z}^{t+1}_i))$, and decode $\mathbf{z}^{t+1}_i=\mathcal{D}(\mathbf{u}^{t+1}_i)$ with decoder $\mathcal{D}$. Note that due to nonlinearities in $\mathcal{E}$ and $\mathcal{D}$, \eqref{eq:diffusion prior step} does not necessarily correspond to a reverse \textit{latent} diffusion with noise level $\rho_{\text{d}}$. A separate schedule for the starting noise level may be needed for diffusion in latent space.

To select batches of slices from a volume, we construct a set cover problem and utilize randomized rounding, followed by a lightweight alteration step that greedily covers any under-covered indices. We follow classic randomized rounding with alteration \cite{raghavan1987randomizedrounding, bansal2012rrwithalteration} which covers every slice at least $r$ times within a specified number of batches. We then perform the likelihood step, the diffusion prior step, and the TV prior step on the selected batch, before averaging the sampled slices that have multiple coverage. This method better preserves stochasticity and explores the volume more uniformly than deterministic methods such as sliding window reconstruction, leading to reduced bias and aliasing.

\begin{algorithm}[t]
\caption{PSI3D}
\label{alg}
\textbf{Input:} initialize $\mathbf{x}^0,\mathbf{z}^0,\mathbf{w}^0$; iterations $T$; forward model $\mathbf{A}$, measurements $\mathbf{y}$; coupling schedules $\{\rho_{\text{d}}\},\{\rho_{\text{tv}}\}$.\\
\textbf{Output:} posterior samples $\{\mathbf{x}^t\}$
\begin{algorithmic}[1]
\For{$t=0, 1, ..., T-1$}
  \State $\mathbf{x}^{t+1} \leftarrow \verb|LikelihoodStep| \big(\mathbf{z}^{t}, \mathbf{w}^{t}, \mathbf{y},\mathbf{A},\rho_{\text{d}},\rho_{\text{tv}}\big)$
  \State $\mathbf{z}^{t+1} \leftarrow \verb|DiffusionPrior| \big( \mathbf{x}^{t+1};\,\rho_{\text{d}}\big)$
  \State $\mathbf{w}^{t+1} \leftarrow \verb|TVPriorStep|\big(\mathbf{x}^{t+1};\rho_{\text{tv}}\big)$
\EndFor
\State \Return $\{\mathbf{x}^t\}$
\end{algorithmic}
\end{algorithm}

\begin{figure}
    \centering
    \includegraphics[width=0.9\linewidth]{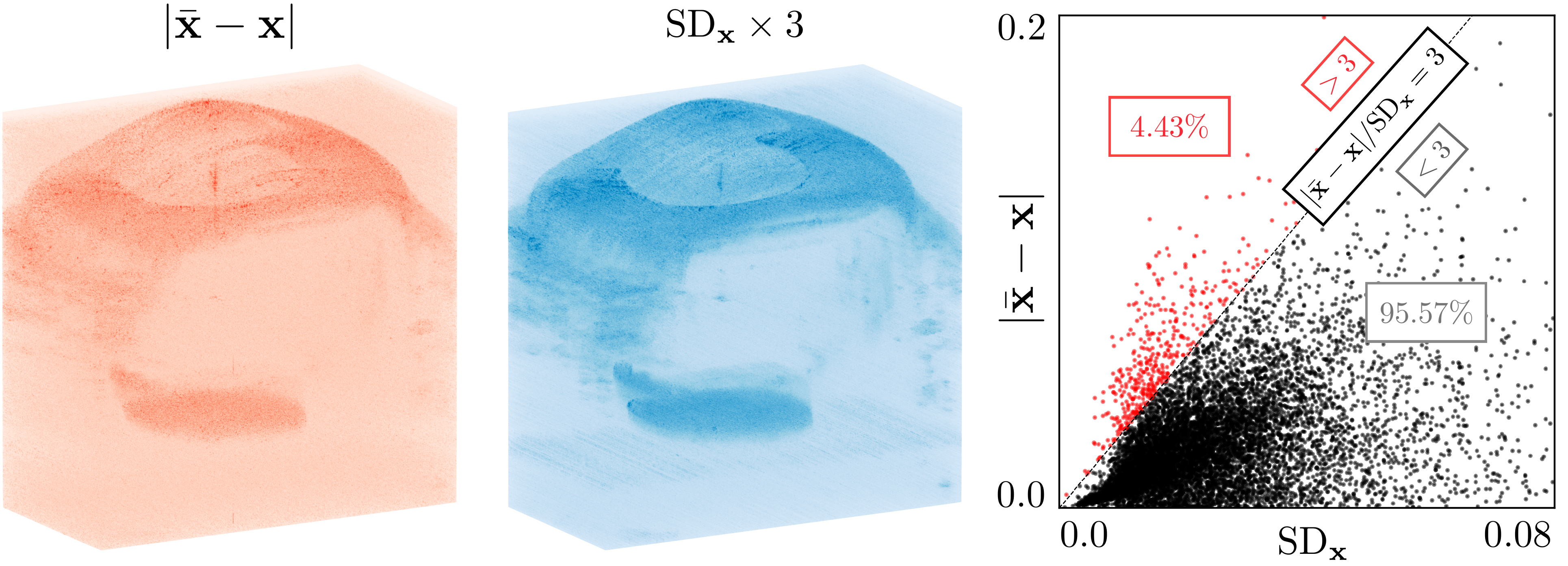}
    \caption{Visualization of pixel-wise statistics associated with the example volume shown in Fig.~\ref{fig:volumes}. We plot the volume's absolute error ($|\bar{\mathbf{x}} - \mathbf{x}|$), $3\times$ standard deviation ($\text{SD}_\mathbf{x}$), and 3-SD credible interval with $10,000$ voxels randomly selected from the volume.}
    \label{fig:uq}
\end{figure}

\section{Experiments and Results}

We validate PSI3D on OCT super-resolution, an application that requires high resolution, minimal hallucinations, and reconstruction credibility to provide proper clinical guidance. Typical OCT volumetric imaging for living tissues takes several seconds, during which even slight motions may cause significant blurring and motion artifacts \cite{yun2004octmotion, baghaie2017octmotion, chhablani2014octartifacts}. 
Recent advances propose to use a high scanning rate to obtain fast but undersampled data, then formulate an inverse problem to reconstruct the original volume \cite{zuo2024octunet, ploner2020octmotioncorrect}. 
Our dataset consists of $164$ OCT volumes with $1024 \times 1024 \times 128$ voxels imaged on fish eyes \cite{zuo2024octunet, wang.etal2025pnpdmoct}. We degrade each B-scan image ($1024 \times 1024$) by $4$ times downsampling and split our dataset into $134$ for training, $15$ for validation, and $15$ for testing. 

For our latent diffusion prior, we first train a VQGAN model on B-scan images to project each slice onto a $64^3$ latent code. We use Gumbel quantization \cite{jang2016gumbel, ramesh2021dall-e} and train for $280,000$ total image iterations without GAN loss for a smoother latent representation. 
We then train an EDM diffusion on the latent codes for $125,000$ steps with a batch size of $64$, while changing data standard deviations to $0.20$ to match that of the latent codes. 
For EDM sampling, we use an annealing schedule for $\rho_{\text{d}}$ that exponentially decays from $5.0$ to $0.025$. Due to encoder-decoder nonlinearities, we determine a separate schedule that decays from $5.0$ to $2.0$ for the diffusion starting noise level. We also determine $S_\text{noise} = 2.2$ heuristically for the best sampling performance. 
We have $\rho_\text{tv}$ decaying from $5.0$ to $2.0$ to match the latent schedule.
Finally, we use $12$ batches of $16$ B-scan slices to stochastically cover a volume of $128$ slices. The likelihood step and the diffusion prior step update each slice before the TV prior step operates inter-slice for each batch.

To measure image reconstruction quality, we employ peak signal-to-noise ratio (PSNR), structural similarity index measure (SSIM), multi-scale structural similarity index measure (MS-SSIM), and perceptual image patch similarity (LPIPS). We compare PSI3D with four baselines: (1) bilinear interpolation, (2) bicubic interpolation, (3) 3D total variation denoising \cite{chambolle2011}, and (4) 3D UNet, where we extend \cite{cciccek2016unet3d} to our scale via patch-wise training and sliding window reconstruction. We train \cite{cciccek2016unet3d} for $1,000$ epochs until convergence on patches of $256 \times 256 \times 16$ randomly cropped from our training volumes. 

We show example reconstructions of two B-scan slices in Fig.~\ref{fig:slices}, along with PSNR and SSIM values for the slices. 
Note how PSI3D accurately reconstructs fine details in the image, especially in the yellow zoomed regions.
For a comprehensive evaluation, we measure our performance using $15$ test volumes. 
Table~\ref{table} shows the reconstruction quality of PSI3D and the four baselines. For PSI3D, we include the statistics for both single posterior samples and the mean $\bar{\mathbf{x}}$, where $\bar{\mathbf{x}}$ averages $20$ samples in $40$ steps after convergence within a single inference chain. We note that single samples from PSI3D may be less competitive than solutions from deterministic solvers, but we can average samples and obtain $\bar{\mathbf{x}}$ in a single inference chain without additional computation. The PSI3D mean $\bar{\mathbf{x}}$ consistently outperforms other methods in PSNR, SSIM, and MS-SSIM, while the PSI3D sample provides the best LPIPS averaged across the slices.

Fig.~\ref{fig:uq} visualizes PSI3D's reconstruction credibility. 
For a single test volume, we use the mean reconstruction to characterize PSI3D's absolute error ($|\bar{\mathbf{x}} - \mathbf{x}|$), 3-standard deviation ($\text{SD}_\mathbf{x} \times 3$), and 3-SD credible interval with $10,000$ voxels randomly selected from the volume. 
PSI3D recovers $95.57\%$ of voxels within the 3-SD interval for this test volume, and $95.55\%$ for all $15$ test volumes. 
We also compute the normalized negative log-likelihood (NLL)\cite{Lakshminarayanan.etal2017} for all test volumes to be $-1.78$ assuming independent pixel-wise Gaussian distributions. 
Our results show that PSI3D yields robust and credible reconstructions from posterior sampling.

\begin{table}[t]
\centering
\caption{Reconstruction quality of PSI3D and other baselines. All numerical results are measured from $15$ test volumes. Note that PSNR, SSIM, and MS-SSIM are evaluated on the 3D volumes, whereas LPIPS is evaluated on 2D B-scan images and averaged across the slices.}
\begin{tabular}{lcccc}
\toprule
\textbf{Method} \rule[-1ex]{0pt}{0pt} & PSNR $\uparrow$ & SSIM $\uparrow$ & MS-SSIM $\uparrow$ & LPIPS\textsuperscript{1} $\downarrow$ \\ \hline
Bilinear \rule{0pt}{3ex} & 27.290 & 0.415 & 0.859 & 0.585 \\
Bicubic & 27.421 & 0.423 & 0.864 & 0.591 \\
TV3D \cite{chambolle2011} & 27.240 & 0.399 & 0.854 & 0.595 \\
UNet3D \cite{cciccek2016unet3d} & 27.968 & 0.482 & 0.903 & 0.493 \\
PSI3D (no TV) & 28.732 & 0.609 & 0.922 & 0.284 \\
PSI3D (sample) & 27.224 & 0.542 & 0.925 & \textbf{0.272} \\
PSI3D (mean) & \textbf{28.990} & \textbf{0.621} & \textbf{0.943} & 0.306 \\
\bottomrule
\end{tabular}
\label{table}
\end{table}

\section{Conclusion}

In this work, we propose PSI3D, a plug-and-play method with latent diffusion and total variation prior for large-scale, stochastic inference. PSI3D consists of a likelihood step, a 2D diffusion prior step, and a 1D TV prior step and iteratively updates between the three to draw posterior samples. We also incorporate stochastic set cover to draw batch samples from the volume with randomness guarantees. PSI3D significantly improves OCT reconstruction quality over traditional and learned baselines, as well as providing robust sampling capabilities.

\bibliographystyle{IEEEtran}
\bibliography{references}

@inproceedings{Brose.etal2017,
	author = {Brosse, Nicolas and Durmus, Alain and Moulines, {\'E}ric and Pereyra, Marcelo},
	booktitle = {Proceedings of Thirtieth Conference on Learning Theory},
	pages = {319--342},
	title = {Sampling from a log-concave distribution with compact support with proximal Langevin Monte Carlo},
	volume = {65},
	year = {2017}}

@article{Pereyra2016,
	author = {Pereyra, Marcelo},
	journal = {Statistics and Computing},
	number = {4},
	pages = {745--760},
	title = {Proximal Markov chain Monte Carlo algorithms},
	volume = {26},
	year = {2016}}

@article{Coeurdoux.etal2024,
  author={Coeurdoux, Florentin and Dobigeon, Nicolas and Chainais, Pierre},
  journal={IEEE Transactions on Image Processing}, 
  title={Plug-and-Play Split Gibbs Sampler: Embedding Deep Generative Priors in Bayesian Inference}, 
  year={2024},
  volume={33},
  pages={3496-3507},
  doi={10.1109/TIP.2024.3404338}
}

@inproceedings{Xu.etal2024provably,
	author = {Xu, Xingyu and Chi, Yuejie},
	booktitle = {Advances in Neural Information Processing Systems},
	pages = {36148--36184},
	title = {Provably Robust Score-Based Diffusion Posterior Sampling for Plug-and-Play Image Reconstruction},
	volume = {37},
	year = {2024}
}

@article{Lakshminarayanan.etal2017,
	author = {Lakshminarayanan, Balaji and Pritzel, Alexander and Blundell, Charles},
	journal = {Advances in neural information processing systems},
	title = {Simple and scalable predictive uncertainty estimation using deep ensembles},
	volume = {30},
	year = {2017}}

@article{Feng.etal2023scorebased,
	author = {Berthy T. Feng and Jamie Smith and Michael Rubinstein and Huiwen Chang and Katherine L. Bouman and William T. Freeman},
	journal = {arXiv:2304.11751 [cs.CV]},
	title = {Score-Based Diffusion Models as Principled Priors for Inverse Imaging},
	year = {2023}}

@article{Bouman.etal2023generative,
	author = {Charles A. Bouman and Gregery T. Buzzard},
	journal = {arXiv:2306.07233 [cs.CV]},
	title = {Generative Plug and Play: Posterior Sampling for Inverse Problems},
	year = {2023}}

@inproceedings{Chung.etal2023diffusion,
	author = {Hyungjin Chung and Jeongsol Kim and Michael Thompson Mccann and Marc Louis Klasky and Jong Chul Ye},
	booktitle = {International Conference on Learning Representations},
	title = {Diffusion Posterior Sampling for General Noisy Inverse Problems},
	year = {2023}}

@inproceedings{Song.etal2019,
	author = {Song, Yang and Ermon, Stefano},
	booktitle = {Advances in Neural Information Processing Systems},
	title = {Generative Modeling by Estimating Gradients of the Data Distribution},
	volume = {32},
	year = {2019}}

@article{Kamilov.etal2023,
	author = {Kamilov, Ulugbek S. and Bouman, Charles A. and Buzzard, Gregery T. and Wohlberg, Brendt},
	journal = {IEEE Signal. Proc. Mag.},
	number = {1},
	pages = {85-97},
	title = {Plug-and-Play Methods for Integrating Physical and Learned Models in Computational Imaging: Theory, algorithms, and applications},
	volume = {40},
	year = {2023}}

@article{Sun.etal2020,
	author = {Sun, Yu and Wu, Zihui and Xu, Xiaojian and Wohlberg, Brendt and Kamilov, Ulugbek S.},
	journal = {IEEE Trans. Comput. Imag.},
	pages = {849-863},
	title = {Scalable Plug-and-Play {ADMM} With Convergence Guarantees},
	volume = {7},
	year = {2021}}

@article{Tian.etal14,
	author = {L. Tian and J. Wang and L. Waller},
	journal = {Opt. Lett.},
	month = {Mar},
	number = {5},
	pages = {1326--1329},
	title = {{3D} differential phase-contrast microscopy with computational illumination using an LED array},
	volume = {39},
	year = {2014}}

@article{Beck.Teboulle2009,
	author = {A. Beck and M. Teboulle},
	journal = {SIAM J. Imaging Sci.},
	number = {1},
	pages = {183--202},
	title = {A Fast Iterative Shrinkage-Thresholding Algorithm for Linear Inverse Problems},
	volume = {2},
	year = {2009}}

@article{Geman.Yang1995,
	author = {Geman, D. and Yang, C.},
	journal = {IEEE Trans. Image Process.},
	month = {July},
	number = {7},
	pages = {932--946},
	title = {Nonlinear Image Recovery with Half-Quadratic Regularization},
	volume = {4},
	year = {1995}}

@conference{Venkatakrishnan.etal2013,
	address = {Austin, TX, USA},
	author = {Venkatakrishnan, S. V. and Bouman, C. A. and Wohlberg, B.},
	booktitle = {Proc. IEEE Global Conf. Signal Process. and Inf. Process. ({GlobalSIP})},
	month = {Dec. 3-5,},
	pages = {945--948},
	title = {Plug-and-Play Priors for Model Based Reconstruction},
	year = {2013}}

@article{Chambolle.2004,
	author = {Chambolle, A.},
	journal = {J. of Math. Imag. and Vis.},
	number = {1},
	pages = {89--97},
	title = {An Algorithm for Total Variation Minimization and Applications},
	volume = {20},
	year = {2004}}

@article{Buzzard.etal2017,
	author = {Buzzard, G. T. and Chan, S. H. and Sreehari, S. and Bouman, C. A.},
	journal = {SIAM J. Imaging Sci.},
	number = {3},
	pages = {2001--2020},
	title = {Plug-and-Play Unplugged: {O}ptimization free reconstruction using consensus equilibrium},
	volume = {11},
	year = {2018}}

@article{Barutcu:2021,
	author = {Barutcu, Semih and Aslan, Selin and Katsaggelos, Aggelos K. and G{\"u}rsoy, Do{\u g}a},
	date = {2021/09/06},
	journal = {Scientific Reports},
	number = {1},
	pages = {17740},
	title = {Limited-angle computed tomography with deep image and physics priors},
	volume = {11},
	year = {2021}}

@article{Ahmad.etal2019,
	author = {R. Ahmad and C. A. Bouman and G. T. Buzzard and S. H. Chan and S.Liu and E.T. Reehorst and P. Schniter},
	journal = {IEEE Signal Process. Mag.},
	month = {Jan.},
	number = {1},
	pages = {105--116},
	title = {Plug-and-Play Methods for Magnetic Resonance Imaging: Using Denoisers for Image Recovery},
	volume = {37},
	year = {2020}}

@inproceedings{Ryu.etal2019,
	author = {E. K. Ryu and J. Liu and S. Wang and X. Chen and Z. Wang and W. Yin},
	booktitle = {Proc. 36th Int. Conf. Machine Learning (ICML)},
	pages = {5546--5557},
	title = {Plug-and-Play Methods Provably Converge with Properly Trained Denoisers},
	volume = {97},
	year = {2019}}

@article{Chan.etal2016,
	author = {Chan, S. H. and Wang, X. and Elgendy, O. A.},
	journal = {IEEE Trans. Comput. Imag.},
	month = {Mar.},
	number = {1},
	pages = {84--98},
	title = {Plug-and-Play {ADMM} for Image Restoration: Fixed-Point Convergence and Applications},
	volume = {3},
	year = {2017}}

@article{chambolle2011,
	author = {Chambolle, Antonin and Pock, Thomas},
	journal = {Journal of mathematical imaging and vision},
	number = {1},
	pages = {120--145},
	publisher = {Springer},
	title = {A first-order primal-dual algorithm for convex problems with applications to imaging},
	volume = {40},
	year = {2011}}

@article{Akiyama.etal2019,
	author = {Akiyama, K. and Alberdi, A. and Alef, W. and Asada, K. and Azulay, R. and Baczko, A.K. and Ball, D. and Balokovi{\'c}, M. and Barrett, J. and Bintley, D. and others},
	journal = {The Astrophysical Journal Letters},
	number = {1},
	pages = {L4},
	title = {First M87 event horizon telescope results. IV. Imaging the central supermassive black hole},
	volume = {875},
	year = {2019}}

@ARTICLE{Sun.etal2024,
  author={Sun, Yu and Wu, Zihui and Chen, Yifan and Feng, Berthy T. and Bouman, Katherine L.},
  journal={IEEE Transactions on Computational Imaging}, 
  title={Provable Probabilistic Imaging Using Score-Based Generative Priors}, 
  year={2024},
  volume={10},
  number={},
  pages={1290-1305},
  doi={10.1109/TCI.2024.3449114}}

@article{wu.etal2024pnpdm,
  title={Principled probabilistic imaging using diffusion models as plug-and-play priors},
  author={Wu, Zihui and Sun, Yu and Chen, Yifan and Zhang, Bingliang and Yue, Yisong and Bouman, Katherine},
  journal={Advances in Neural Information Processing Systems},
  volume={37},
  pages={118389--118427},
  year={2024}
}

@article{karras2022edm,
  title={Elucidating the design space of diffusion-based generative models},
  author={Karras, Tero and Aittala, Miika and Aila, Timo and Laine, Samuli},
  journal={Advances in neural information processing systems},
  volume={35},
  pages={26565--26577},
  year={2022}
}

@inproceedings{chung2023diffusionmbir,
  title={Solving 3d inverse problems using pre-trained 2d diffusion models},
  author={Chung, Hyungjin and Ryu, Dohoon and McCann, Michael T and Klasky, Marc L and Ye, Jong Chul},
  booktitle={Proceedings of the IEEE/CVF conference on computer vision and pattern recognition},
  pages={22542--22551},
  year={2023}
}

@article{ozaki2024iterative,
  title={Iterative CT Reconstruction via Latent Variable Optimization of Shallow Diffusion Models},
  author={Ozaki, Sho and Kaji, Shizuo and Imae, Toshikazu and Nawa, Kanabu and Yamashita, Hideomi and Nakagawa, Keiichi},
  journal={arXiv preprint arXiv:2408.03156},
  year={2024}
}

@inproceedings{esser2021taming,
  title={Taming transformers for high-resolution image synthesis},
  author={Esser, Patrick and Rombach, Robin and Ommer, Bjorn},
  booktitle={Proceedings of the IEEE/CVF conference on computer vision and pattern recognition},
  pages={12873--12883},
  year={2021}
}

@article{majee2021mace,
  title={Multi-slice fusion for sparse-view and limited-angle 4D CT reconstruction},
  author={Majee, Soumendu and Balke, Thilo and Kemp, Craig AJ and Buzzard, Gregery T and Bouman, Charles A},
  journal={IEEE Transactions on Computational Imaging},
  volume={7},
  pages={448--462},
  year={2021},
  publisher={IEEE}
}

@article{raghavan1987randomizedrounding,
  title={Randomized rounding: a technique for provably good algorithms and algorithmic proofs},
  author={Raghavan, Prabhakar and Tompson, Clark D},
  journal={Combinatorica},
  volume={7},
  number={4},
  pages={365--374},
  year={1987},
  publisher={Springer}
}

@article{bansal2012rrwithalteration,
  title={Solving packing integer programs via randomized rounding with alterations},
  author={Bansal, Nikhil and Korula, Nitish and Nagarajan, Viswanath and Srinivasan, Aravind},
  journal={Theory of Computing},
  volume={8},
  number={24},
  pages={533--565},
  year={2012},
  publisher={University of Chicago Press}
}

@article{yun2004octmotion,
  title={Motion artifacts in optical coherence tomography with frequency-domain ranging},
  author={Yun, SH and Tearney, GJ and De Boer, JF and Bouma, BE},
  journal={Optics Express},
  volume={12},
  number={13},
  pages={2977--2998},
  year={2004},
  publisher={Optical Society of America}
}

@article{baghaie2017octmotion,
  title={Involuntary eye motion correction in retinal optical coherence tomography: Hardware or software solution?},
  author={Baghaie, Ahmadreza and Yu, Zeyun and D’Souza, Roshan M},
  journal={Medical image analysis},
  volume={37},
  pages={129--145},
  year={2017},
  publisher={Elsevier}
}

@article{chhablani2014octartifacts,
  title={Artifacts in optical coherence tomography},
  author={Chhablani, Jay and Krishnan, Tandava and Sethi, Vaibhav and Kozak, Igor},
  journal={Saudi Journal of Ophthalmology},
  volume={28},
  number={2},
  pages={81--87},
  year={2014},
  publisher={Elsevier}
}

@article{zuo2024octunet,
  title={High-resolution in vivo 4D-OCT fish-eye imaging using 3D-UNet with multi-level residue decoder},
  author={Zuo, Ruizhi and Wei, Shuwen and Wang, Yaning and Irsch, Kristina and Kang, Jin U},
  journal={Biomedical optics express},
  volume={15},
  number={9},
  pages={5533--5546},
  year={2024},
  publisher={Optica Publishing Group}
}

@article{ploner2020octmotioncorrect,
  title={Efficient and high accuracy 3-D OCT angiography motion correction in pathology},
  author={Ploner, Stefan B and Kraus, Martin F and Moult, Eric M and Husvogt, Lennart and Schottenhamml, Julia and Yasin Alibhai, A and Waheed, Nadia K and Duker, Jay S and Fujimoto, James G and Maier, Andreas K},
  journal={Biomedical Optics Express},
  volume={12},
  number={1},
  pages={125--146},
  year={2020},
  publisher={Optical Society of America}
}

@article{jang2016gumbel,
  title={Categorical reparameterization with gumbel-softmax},
  author={Jang, Eric and Gu, Shixiang and Poole, Ben},
  journal={arXiv preprint arXiv:1611.01144},
  year={2016}
}

@inproceedings{ramesh2021dall-e,
  title={Zero-shot text-to-image generation},
  author={Ramesh, Aditya and Pavlov, Mikhail and Goh, Gabriel and Gray, Scott and Voss, Chelsea and Radford, Alec and Chen, Mark and Sutskever, Ilya},
  booktitle={International conference on machine learning},
  pages={8821--8831},
  year={2021},
  organization={Pmlr}
}

@inproceedings{cciccek2016unet3d,
  title={3D U-Net: learning dense volumetric segmentation from sparse annotation},
  author={{\c{C}}i{\c{c}}ek, {\"O}zg{\"u}n and Abdulkadir, Ahmed and Lienkamp, Soeren S and Brox, Thomas and Ronneberger, Olaf},
  booktitle={International conference on medical image computing and computer-assisted intervention},
  pages={424--432},
  year={2016},
  organization={Springer}
}

@article{vono.dobigeon.chainais2022,
  title={High-dimensional Gaussian sampling: A review and a unifying approach based on a stochastic proximal point algorithm},
  author={Vono, Maxime and Dobigeon, Nicolas and Chainais, Pierre},
  journal={SIAM Review},
  volume={64},
  number={1},
  pages={3--56},
  year={2022},
  publisher={SIAM}
}

@article{vono.dobigeon.chainais2019,
  title={Split-and-augmented Gibbs sampler—Application to large-scale inference problems},
  author={Vono, Maxime and Dobigeon, Nicolas and Chainais, Pierre},
  journal={IEEE Transactions on Signal Processing},
  volume={67},
  number={6},
  pages={1648--1661},
  year={2019},
  publisher={IEEE}
}

@article{wang.etal2025pnpdmoct,
  title={Super-Resolution Optical Coherence Tomography Using Diffusion Model-Based Plug-and-Play Priors},
  author={Wang, Yaning and Yu, Jinglun and Guo, Wenhan and Sun, Yu and Kang, Jin U},
  journal={arXiv preprint arXiv:2505.14916},
  year={2025}
}

\end{document}